\documentclass[12pt,a4paper,final]{iopart}
\usepackage{iopams}  
\usepackage{graphicx}
\usepackage{hyperref}
\usepackage{cleveref}
\usepackage[bottom]{footmisc}
\begin{document}
\title[Einstein Flow]{Einstein flow with matter sources: stability and convergence}
\author{Vincent Moncrief$^{1,2}$\footnote{e-mail: vincent.moncrief@yale.edu}, Puskar Mondal$^{3}$ \footnote{e-mail: puskar@cmsa.fas.harvard.edu}}

\begin{abstract}
Two recent articles \cite{ashtekar2015general, moncrief2019could} suggested an interesting dynamical mechanism within the framework of the vacuum Einstein flow (or Einstein-$\Lambda$ flow if a positive cosmological constant $\Lambda$ is included) which suggests that many closed (compact without boundary) manifolds that do not support homogeneous and isotropic metrics \textit{at all} will nevertheless evolve to be asymptotically compatible with the observed approximate homogeneity and isotropy of the physical universe. These studies however did not include matter sources. Therefore the aim of the present study is to include suitable matter sources and investigate whether one is able to draw a similar conclusion. 
\end{abstract}
 
\section{Introduction}     
Viewed on a sufficiently coarse-grained scale the portion of our universe that is
accessible to observation appears to be spatially homogeneous and isotropic.
If, as is usually imagined, one should be able to extrapolate these features
to (a suitably coarse-grained model of) the universe as a whole then only a
handful of spatial manifolds need be considered in cosmology — the familiar
Friedmann-Lemaître-Robertson-Walker (FLRW) archetypes of constant
positive, vanishing or negative curvature \cite{ashtekar2015general, cosmo}. These geometries consist,
up to an overall, time-dependent scale factor, of the 3-sphere, $\mathbb{S}^{3}$, with its
canonical ‘round’ metric, Euclidean 3-space, $\mathbb{E}^{3}$, hyperbolic 3-space, $\mathbb{H}^{3}$ and
the quotient space $\mathbb{RP}(3)= \mathbb{S}^{3}/\mathbb{Z}_{2}$ obtainable from $\mathbb{S}^{3}$ by the identification
of antipodal points \cite{wolf}. Of these possibilities only the sphere and its 2-fold
quotient $\mathbb{RP}^{3}$ are closed and thus compatible with a universe model of finite
extent. It is not known of course whether the actual universe is spatially
closed or not but, to simplify the present discussion, we shall limit our attention
herein to models that are. More precisely we shall focus on spacetimes
admitting Cauchy hypersurfaces that are each diffeomorphic to a smooth,
connected 3-manifold that is compact, orientable and without boundary.

On the other hand if one takes literally the cosmological principle that
only manifolds supporting a globally homogeneous and isotropic metric should
be considered in models for the actual universe then, within the spatially
compact setting considered here, only the 3-sphere and $\mathbb{RP}^{3}$ would remain.
But the astronomical observations which motivate this principle are necessarily
limited to a (possibly quite small) fraction of the entire universe and
are compatible with models admitting metrics that are only locally, but not
necessarily globally, spatially homogeneous and isotropic. As is well-known
there are spatially compact variants of all of the basic Friedmann-Lemaître-
Robertson-Walker cosmological models, mathematically constructable (in the
cases of vanishing or negative curvature) by taking suitable compact quotients
of Euclidean 3-space $\mathbb{E}^{3}$ or of hyperbolic 3-space $\mathbb{H}^{3}$. One can also take infinitely
many possible quotients of $\mathbb{S}^{3}$ to obtain the so-called spherical space
forms that are locally compatible with the FLRW constant positive curvature
geometry but are no longer diffeomorphic to the 3-sphere.

Still more generally though we shall find that there is a dynamical mechanism
at work within the Einstein ‘flow’, suitably viewed in terms of the evolution
of 3-manifolds to develop 4-dimensional, globally hyperbolic spacetimes,
and extended to include suitable matter sources and a positive cosmological constant $\Lambda$, that strongly
suggests that even manifolds that do not admit a locally homogeneous and
isotropic metric at all will nevertheless evolve in such a way as to be asymptotically
compatible with the observed homogeniety and isotropy. This reflects
an argument which we shall sketch that, under Einsteinian evolution, the
summands making up $M$ (in a connected sum decomposition) that do support
locally homogeneous and isotropic metrics will tend to overwhelmingly dominate the spatial volume asymptotically as the universe model continues to expand and furthermore that the actual evolving (inhomogeneous, non-isotropic) metric on M will naturally tend to flow towards a homogeneous, isotropic one on each of these asymptotically volume-dominating summands.

We do not claim that this mechanism is yet so compelling, either mathematically
or physically, as to convince one that the actual universe has a more exotic topology but only that such a possibility is not strictly excluded
by current observations. However, it is intriguing to investigate the possibility that there may be a dynamical reason, provided by Einstein’s equations, for the observed fact that the universe seems to be at least locally homogeneous
and isotropic and that this mechanism may therefore allow an attractive logical
alternative to simply extrapolating observations of necessarily limited scope to the universe as a whole.

But what are the (compact, connected, orientable) 3-manifolds available for consideration? This question has been profoundly clarified in recent years by the dramatic progress on lower dimensional topology made possible through the advancements in Ricci flow \cite{ricci, morgan2007ricci}. One now knows for example that,
since the Poincaré conjecture has finally been proven, any such 3-manifold
M that is in fact simply connected must be diffeomorphic to the ordinary
3-sphere $\mathbb{S}^{3}$. Setting aside this so-called ‘trivial’ manifold the remaining possibilities
consist of an infinite list of nontrivial manifolds, each of which is
diffeomorphic (designated herein by $\approx$) to a finite connected sum of the following
form:
\begin{eqnarray}
\label{eq:cs}
M\approx\\\nonumber
\mathbf{S}^{3}/\Gamma_{1}\#..\#\mathbf{S}^{3}/\Gamma_{k}\#(\mathbf{S}^{2}\times \mathbf{S}^{1})_{1}\#..\#(\mathbf{S}^{2}\times \mathbf{S}^{1})_{l}\#\mathbf{K}(\pi,1)_{1}\#..\#\mathbf{K}(\pi,1)_{m}.
\end{eqnarray}
Here $k,l$ and $m$ are integers $\geq 0, k+l+m \geq 1$ and if either $k$, $l$ or $m$ is 0 then
terms of that type do not occur. The connected sum $M\#N$ of two closed
connected, oriented n-manifolds is constructed by removing the interiors of
an embedded closed n-ball in each of $M$ and $N$ and then identifying the resulting
$\mathbb{S}^{n-1}$ boundary components by an orientation-reversing diffeomorphism.
The resulting $n-$manifold will be smooth, connected, closed and consistently
oriented with the original orientations of $M$ and $N$. The above decomposition
of $M$ is only uniquely defined provided we set aside $\mathbb{S}^{3}$ since $M\# \mathbb{S}^{3}\approx M$ for any 3-manifold $M$.

In the above formula if $k\geq 1$, then each $\Gamma_{i}$, $1 \leq i \leq k$ is a finite, nontrivial
($\Gamma_{i}\neq I$) subgroup of SO(4) acting freely and orthogonally on $\mathbb{S}^{3}$. The individual summands $\mathbb{S}^{3}/\Gamma_{i}$ are the spherical space forms alluded to previously
and, by construction, each is compatible with an FLRW metric of constant
positive spatial curvature (i.e., $k = +1$ models in the usual notation). The
individual ‘handle’ summands $\mathbb{S}^{2}\times \mathbb{S}^{1}$ admit metrics of the Kantowski-Sachs
type that are homogeneous but not isotropic and so not even locally of FLRW
type.

The remaining summands in the above ‘prime decomposition’ theorem \cite{prime} are the $K(\pi, 1)$ manifolds of Eilenberg-MacLane type wherein, by definition $\pi = \pi_{1}(M)$, the fundamental group of $M$ and all of the higher
homotopy groups are trivial, that is $\pi_{i}(M)=0$ for $i>1$. Equivalently, the
universal covering space of $M$ is contractible and, in this case, known to be
diffeomorphic to $\mathbb{R}^{3}$ \cite{prime}. Since the higher homotopy groups, $\pi_{i}(M)$ for $i>1$ can be interpreted as the homotopy classes of continuous maps $\mathbb{S}^{i}\to M$,
each such map must be homotopic to a constant map. For this reason $K(\pi,1)$
manifolds are said to be aspherical.

This general class of $K(\pi,1)$ manifolds includes, as special cases, the
3-torus and five additional manifolds, finitely covered by the torus, that are
said to be of ‘flat type’ since they are the only compact, connected, orientable
3-manifolds that each, individually, admits a flat metric and thus supports
spatially compactified versions of the FLRW spaces of flat type (i.e., $k=0$ models).

Other $K(\pi,1)$ spaces include the vast set of compact hyperbolic manifolds
$\mathbb{H}^{3}/\Gamma$, where here $\Gamma$ is a discrete torsion-free (i.e., no nontrivial element has finite order) co-compact subgroup of the Lie group Isom$^{+}$($\mathbb{H}^{3}$) of orientation preserving
isometries of $\mathbb{H}^{3}$ that, in fact, is Lie-group isomorphic to the proper
orthochronous Lorentz group SO$^{+}$(3, 1). Each of these, individually, supports
spatially compactified versions of the FLRW spacetimes of constant negative
(spatial) curvature (i.e., $k = -1$ models).

Additional $K(\pi,1)$ manifolds include the trivial circle bundles over higher
genus surfaces $\Sigma_{p}$ for $p\geq 2$ (where $\Sigma_{p}$ designates a compact, connected, orientable
surface of genus $p$) and nontrivial circle bundles over $\Sigma_{p}$ for $p\geq 1$.
Note that the trivial circle bundles $\mathbb{S}^{2}\times \mathbb{S}^{1}$ and $\mathbb{T}^{2}\times \mathbb{S}^{1} \approx \mathbb{T}^{3}$ are already
included among the previous prime factors discussed and that nontrivial circle
bundles over $\mathbb{S}^{2}$ are included among the spherical space forms $\mathbb{S}^{3}/\Gamma$ for
suitable choices of $\Gamma$. Still further examples of $K(\pi,1)$ manifolds are compact 3-manifolds that
fiber nontrivially over the circle with fiber $\Sigma_{p}$ for $p\geq 1$. Any such manifold
is obtained by identifying the boundary components of $[0, 1]\times \Sigma_{p}$ with a
(nontrivial) orientation-reversing diffeomorphism of $\Sigma_{p}$.

It is known however that every prime $K(\pi,1)$ manifold is decomposible
into a (possibly trivial but always finite) collection of (complete, finite volume)
hyperbolic and graph manifold components. The possibility of such a
(nontrivial) decomposition arises whenever the $K(\pi,1)$ manifold under study
admits a nonempty family $\{T_{i}\}$ of disjoint embedded incompressible two-tori.
An embedded two-torus $\mathbb{T}^{2}$ is said to be incompressible if every incontractible
loop in the torus remains incontractible when viewed as a loop in the ambient
manifold. A closed oriented 3-manifold G (possibly with boundary) is
a graph manifold if there exists a finite collection $\{T^{'}_{i}
\}$ of disjoint embedded
incompressible tori $\{T^{'}_{i}
\}\subset$G such that each component $G_{j}$ of $G-\cup T^{'}_{i}$ is a Seifert-fibered space (A Seifert-fibered space is a 3-manifold foliated by circular fibers in such a way
that each fiber has a tubular neighborhood (characterized by a pair of co-prime
integers) of the special type known as a standard fibered torus).  Thus a graph manifold is a union of Seifert-fibered spaces glued together by toral automorphisms along toral boundary components.
The collection of tori is allowed to be empty so that, in particular, a
Seifert-fibered manifold itself is a graph manifold. Decomposing a 3-manifold
by cutting along essential two-spheres (to yield its prime factors) and then
along incompressible tori, when present, are the basic operations that reduce
a manifold to its ‘geometric’ constituents \cite{prime}. The Thurston conjecture that
every such 3-manifold can be reduced in this way has now been established
via arguments employing Ricci flow \cite{ricci, morgan2007ricci}.

It may seem entirely academic to consider such general, ‘exotic’ 3-manifolds
as the composite (i.e., nontrivial connected sum) ones described above
as arenas for general relativity when essentially all of the explicitly known
solutions of Einstein’s equations (in this spatially compact setting) involve
only individual, ‘prime factors’. As we shall see however some rather general
conclusions are derivable concerning the behaviors of solutions to the field
equations on such \textit{exotic} manifolds and astronomical observations do not logically
exclude the possibility that the actual universe could have such a global
topological structure. It is furthermore conceivable that the validity of central
open issues in general relativity like the \textit{cosmic censorship} conjecture could
depend crucially upon the spatial topology of the spacetime under study.

\section{Field equations}         
In our study, the spacetime is described by an `n+1' dimensional globally hyperbolic Lorentzian manifold $\tilde{M}\approx$ $\mathbb{R}\times M$ with each level set $\{t\}\times M$ of the time function $t$ being an orientable n-manifold diffeomorphic to a Cauchy hypersurface and equipped with a Riemannian metric. Throughout our analysis, we will consider $M$ to be of negative Yamabe type (see appendix for detailed description) since this will suffice to ensure that an expanding universe model can never achieve a maximal hypersurface and then `recollapse'. Such a split $\mathbb{R}\times M$ may be executed by introducing a lapse function $N$ and shift vector field $X$ belonging to suitable function spaces and defined such that
\begin{eqnarray}
\partial_{t}&=&N\hat{n}+X
\end{eqnarray}
with $t$ and $\hat{n}$ being the time function and a hypersurface orthogonal future directed timelike unit vector, respectively (so that in particular $\hat{g}(\hat{n},\hat{n})=-1$). The above splitting expresses the spacetime metric $\tilde{g}$ in local coordinates $\{x^{\alpha}\}_{\alpha=0}^{n}=\{t,x^{1},x^{2},....,x^{n}\}$ as 
\begin{eqnarray}
\hat{g}&=&-N^{2}dt\otimes dt+g_{ij}(dx^{i}+X^{i}dt)\otimes(dx^{j}+X^{j}dt)
\end{eqnarray} 
and the stress-energy tensor as
\begin{eqnarray}
\mathbf{T}&=&E\mathbf{n}\otimes\mathbf{n}+\mathbf{J}\otimes \mathbf{n}+\mathbf{n}\otimes \mathbf{J}+\mathbf{S},
\end{eqnarray}
where $\mathbf{J}\in \mathfrak{X}(M)$, $\mathbf{S}\in S^{2}(M)$. Here, $\mathfrak{X}(M)$ and $S^{2}(M)$ are the space of vector fields and the space of symmetric covariant 2-tensors, respectively. Under such an $n+1$ decomposition, the Einstein equations, 
\begin{eqnarray}
R_{\mu\nu}-\frac{1}{2}Rg_{\mu\nu}+\Lambda g_{\mu\nu}&=&T_{\mu\nu},
\end{eqnarray}
take the form (in units for which $8\pi G=c=1$) of evolution equations,
\begin{eqnarray}
\label{eq:feq1}
\partial_{t}g_{ij}&=&-2Nk_{ij}+L_{X}g_{ij},\\\nonumber
\partial_{t}k_{ij}&=&-\nabla_{i}\nabla_{j}N+N\{R_{ij}+\tau k_{ij}-2k_{ik}k^{k}_{j}\\
\label{eq:feq2}
&&-\frac{1}{n-1}(2\Lambda-S+E)g_{ij}-S_{ij}\}+L_{X}k_{ij},
\end{eqnarray}
along with constraints (Gauss and Codazzi equations)
\begin{eqnarray}
\label{eq:HC}
R(g)-|k|^{2}+\tau^{2}&=&2\Lambda+2E,\\
\label{eq:MC1}
\nabla_{j}k^{j}_{i}-\nabla_{i}\tau&=&-J_{i},
\end{eqnarray} 
where $S=g^{ij}S_{ij}$. The vanishing of the covariant divergence of the stress energy tensor i.e., $\nabla_{\nu}T^{\mu\nu}=0$ is equivalent to the continuity equation and equations of motions of the matter
\begin{eqnarray}
\frac{\partial E}{\partial t}&=&L_{X}E+NE\tau-N\nabla_{i}J^{i}-2J^{i}\nabla_{i}N+NS^{ij}k_{ij},\\
\frac{\partial J^{i}}{\partial t}&=&L_{X}J^{i}+N\tau J^{i}-\nabla_{j}(NS^{ij})+2Nk^{i}_{j}J^{j}-E\nabla^{i}N.\nonumber
\end{eqnarray}
Here we impose `constant mean extrinsic curvature' (CMC) as the temporal gauge condition which yields an elliptic equation for the lapse function (notice that the lapse and shift do not have evolution equations; a consequence of the diffeomorphism invariance of the Einstein equations). Later on, we will select a suitable, complementary spatial gauge. CMC gauge entails 
\begin{eqnarray}
\tau=tr_{g}k=monotonic~function~of~t~alone
\end{eqnarray}      
so $\tau$ is thus constant throughout each $t=constant$ hypersurface and therefore can play the role of time. Using the evolution and constraint equations, one may derive the following equation for the lapse function
\begin{eqnarray}
\frac{\partial \tau}{\partial t}&=&\Delta_{g}N+\{|k|^{2}+\frac{S}{n-1}+\frac{n-2}{n-1}E-\frac{2\Lambda}{n-1}\}N+L_{X}\tau
\end{eqnarray}
which after implementing CMC gauge ($\partial_{i} \tau=0$) yields
\begin{eqnarray}
\label{eq:lase}
\frac{\partial \tau}{\partial t}&=&\Delta_{g}N+\{|k|^{2}+\frac{S}{n-1}+\frac{n-2}{n-1}E-\frac{2\Lambda}{n-1}\}N,\nonumber
\end{eqnarray}
where $|k|^{2}=k_{ij}k^{ij}$ and the Laplacian is defined as $\Delta_{g}=-\nabla[g]^{i}\nabla[g]_{i}$ and therefore has positive spectrum on compact connected manifolds. 

Utilizing the `CMC' condition, the momentum constraint (\ref{eq:MC1}) simplifies to 
\begin{eqnarray}
\label{eq:MC}
\nabla[g]_{j}k^{j}_{i}&=&-J_{i},
\end{eqnarray}
the solution of which may be written as 
\begin{eqnarray}
\label{eq:2ndf}
k^{i}_{j}=K^{tri}_{j}+\frac{\tau}{n}\delta^{i}_{j}
\end{eqnarray}
with $K^{tr}$ being traceless with respect to $g$ i.e., $K^{tr}_{ij}g^{ij}=0$ (and so with respect to any metric conformal to $g$). Note that $K^{tr}$ is obtained by solving the following equation
\begin{eqnarray}
\nabla[g]_{j}K^{trj}_{i}&=&-J_{i}.
\end{eqnarray}
The Hamiltonian constraint (\ref{eq:HC}) may be written using the solution of the momentum constraint (\ref{eq:MC}) as follows
\begin{eqnarray}
R(g)&=&|K^{tr}|^{2}+2E-\frac{n-1}{n}(\tau^{2}-\frac{2n\Lambda}{n-1})\nonumber.
\end{eqnarray}
Since we are primarily interested in the case $\Lambda>0$ (the simplest model for `dark energy'), it might appear that $\tau^{2}-\frac{2n\Lambda}{n-1}$ could be negative. But then the Hamiltonian constraint would imply that if the energy density $E$ is non-negative, then $R(g)\geq0$ everywhere on $M$ which is impossible for a manifold of negative Yamabe type (see appendix for details about Yamabe classification). Let's consider the energy condition and establish an allowed range for the constant extrinsic mean curvature $\tau$ which is now playing the role of time. The weak energy condition yields  
\begin{eqnarray}
\mathbf{T}(\mathbf{n},\mathbf{n})\geq0,~E\geq0
\end{eqnarray}
so that to any time-like observer the energy density is nonnegative. Physically relevant classical matter sources are expected to satisfy this energy condition. We will only consider matter sources with point-wise nonnegative energy density throughout the spacetime. Therefore a universe filled with matter satisfying the weak energy condition will always have 
\begin{eqnarray}
\tau^{2}-\frac{2n\Lambda}{n-1}>0,
\end{eqnarray}
and for expanding models (so that $\frac{\partial \tau}{\partial t}>0$),
\begin{eqnarray}
-\infty<\tau<-\sqrt{\frac{2n\Lambda}{n-1}}.
\end{eqnarray} 
Since we are primarily interested in the asymptotic behavior of an `\textbf{expanding}' universe model (the physically relevant case), we set the range of $\tau$ to be $(-\infty, -\sqrt{\frac{2n\Lambda}{n-1}})$ once and for all. We turn our attention to the lapse equation in an expanding universe model by setting (to specify the time gauge precisely) 
\begin{eqnarray}
\label{eq:timefunction}
\frac{\partial \tau}{\partial t}&=&\frac{n}{2(n-1)}(\tau^{2}-\frac{2n\Lambda}{n-1})^{\frac{n}{2}}>0
\end{eqnarray}
whose solution $\tau=\tau(t)$ plays the role of time from now onwards. The lapse equation for this choice of time co-ordinate is expressible as
\begin{eqnarray}
\label{eq:lapse}
\Delta_{g}N+(|K^{tr}|^{2}+(\frac{\tau^{2}}{n}-\frac{2\Lambda}{n-1})+\frac{1}{n-1}(S+(n-2)E))N\\\nonumber 
=\frac{n}{2(n-1)}(\tau^{2}-\frac{2n\Lambda}{n-1})^{\frac{n}{2}}.\nonumber
\end{eqnarray}
a unique solution of which is implied by 
\begin{eqnarray}
\label{eq:energy1}
(n-2)E+S\geq0.
\end{eqnarray}
Note that matter satisfying the strong energy condition   
\begin{eqnarray}
(T_{\mu\nu}-\frac{1}{2}Tg_{\mu\nu})n^{\mu}n^{\nu}\geq0,\\\nonumber
E+S\geq0
\end{eqnarray}
automatically satisfies the condition (\ref{eq:energy1}) for $n\geq 3$ since we are considering $E\geq0$. In a sense the matter fields satisfying both weak and strong energy conditions form a subset in the space of mater fields satisfying $E\geq0$ and $S+(n-2)E\geq0$ for $n\geq 3$ and therefore such sources are allowed for our analysis. Of course, most physically relevant sources do satisfy the property of non-negative energy density (weak energy condition) and the attractive nature of gravity (strong energy condition). Several known sources of physical interest satisfy both the weak and strong energy conditions. These include for example perfect fluids and Vlasov matter (see \cite {rendall1999dynamics, lee2005einstein, andreasson2011einstein, ringstrom2013topology, fajman2019kantowski} for details about Vlasov matter). 

We have established the existence of a unique solution of the lapse equation provided that the matter sources in the universe satisfy a suitable energy condition. A standard maximum principle argument for the corresponding elliptic equation yields the following estimate for the lapse function 
\begin{eqnarray}
\label{eq:NES}
0&<&\frac{\frac{n}{2(n-1)}(\tau^{2}-\frac{2n\Lambda}{n-1})^{\frac{n}{2}}}{\left((\frac{\tau^{2}}{n}-\frac{2\Lambda}{n-1})+\sup(|K^{tr}|^{2}+\frac{1}{n-1}(S+(n-2)E))\right)}\leq N\leq\\\nonumber
&& \frac{n^{2}}{2(n-1)}(\tau^{2}-\frac{2n\Lambda}{n-1})^{\frac{n}{2}-1}.\nonumber
\end{eqnarray}
The important thing here is to note that the lapse function is positive for an expanding universe model. The results obtained so far will be sufficient to study the dynamical behavior in terms of a \textit{weak} Lyapunov functional.

\section{A weak Lyapunov functional and its monotonic decay}
A Lyapunov functional is indispensable while studying the stability properties of a dynamical system. Construction of a Lyapunov functional is however a challenging task. In order to study the dynamical properties of the Einstein-matter system (possibly including a positive cosmological constant), we require a Lyapunov functional for this field theoretic problem. Using a conformal technique, \cite {fischer1997hamiltonian, moncrief2019could} constructed a reduced phase space as the cotangent bundle of the higher dimensional analogue of the  Teichm\"uller space of $2+1$ dimensional gravity and obtained the following true Hamiltonian of the dynamics\footnote{~wherein $g$ is expressed in terms of a `conformal' metric $\gamma$ and a conformal factor $\varphi$ that satisfies the associated Lichnerowicz equation}
\begin{eqnarray}
H_{reduced}&:=&\frac{2(n-1)}{n}\int_{M}\frac{\partial \tau}{\partial t}\mu_{g}.
\end{eqnarray}
In the particular case of a vacuum limit, it is indeed possible to construct such a reduced Hamiltonian in `CMC' gauge, which also acts as a Lyapunov functional. Motivated by their study, we consider the rescaled volume functional as a Lyapunov functional and call it $L(g,k)$ 
\begin{eqnarray}
\label{eq:LF}
L(g,k)&=&\frac{2(n-1)}{n}\int_{M}\frac{\partial \tau}{\partial t}\mu_{g},
\end{eqnarray}
where $g$ is expressed as $g_{ij}=\varphi^{4/n-2}\gamma_{ij},~\varphi:M\to \mathbb{R}_{>0},~R(\gamma)=-1$. 
We call $L(g,k)$ a \textit{weak} Lyapunov functional because, following the expression of $\frac{\partial\tau}{\partial t}$ and the corresponding Lichnerowicz equation for $\varphi$, it controls only the $H^{1}\times L^{2}$ norm of the data $(g,k)$ while the desired norm would more likely be $H^{s}\times H^{s-1}, s>\frac{n}{2}+\epsilon$ for some $\epsilon\geq1$. Note that we do not, at this point, have a local existence theorem for the Cauchy problem of the Einstein system with arbitrary matter sources satisfying the desired energy conditions. At the moment, let us focus on the time evolution of this weak Lyapunov functional. Due to its weak character, we will not be able to state a theorem concerning the stability (either Lyapunov or asymptotic) of the spacetime on the basis of its time evolution. Nevertheless, we will be able to obtain some important physical results related to the asymptotic behavior of the spacetime (in the expanding direction) utilizing the limiting behavior of the Lyapunov function. The time derivative of this functional may be obtained as  
\begin{eqnarray}
\frac{dL(g,k)}{dt}\\
=n\tau(\tau^{2}-\frac{2n\Lambda}{n-1})^{\frac{n}{2}-1}\int_{M}(|K^{tr}|^{2}+\frac{1}{n-1}(S+(n-2)E))N\mu_{g},
\end{eqnarray} 
where we have used the field equations (\ref{eq:feq1}-\ref{eq:feq2}) and the identity
\begin{eqnarray}
\frac{\partial^{2} \tau}{\partial t^{2}}
&=&\frac{n}{2(n-1)}\tau(\tau^{2}-\frac{2n\Lambda}{n-1})^{\frac{n}{2}-1}\left(n\Delta_{g}N+n|K^{tr}|^{2}N\right.\\
&&\left.+(\tau^{2}-\frac{2n\Lambda}{n-1})N+\frac{n}{n-1}(S+(n-2)E)N\right),\nonumber
\end{eqnarray}
which was obtained using the lapse equation and the CMC gauge condition ($\partial_{i} \tau=0$). We have also used Stokes' theorem to eliminate the covariant divergence terms in the integral over the compact manifold $M$. Along the solution curve in the expanding direction ($\frac{\partial \tau}{\partial t}>0$ and $-\infty<\tau<-\sqrt{\frac{2n\Lambda}{n-1}}$), therefore, the Lyapunov functional monotonically decays i.e.,
\begin{eqnarray}
\frac{d L(g,k)}{dt}<0
\end{eqnarray}
and can only attain its infimum (i.e., $\frac{dL}{dt}=0$) if the following conditions are met (perhaps only asymptotically)  
\begin{eqnarray}
K^{tr}&=&0,~S+(n-2)E=0.
\end{eqnarray} 
Substitution of the first condition into the momentum constraint (\ref{eq:MC}) immediately yields 
\begin{eqnarray}
J^{i}\equiv0
\end{eqnarray}
everywhere on $M$. In addition to satisfying $E, S+(n-2)E\geq0$, if the matter sources also satisfy the physically relevant strong energy condition i.e., $(S+E)\geq0$, then for $n\geq3$ (the cases of primary interest), the second condition for the infimum of the Lyapunov function translates to
\begin{eqnarray}
S+(n-2)E=0,~E=0
\end{eqnarray}
and therefore 
\begin{eqnarray}
S=0.
\end{eqnarray}
This result, therefore, states that the weak Lyapunov functional is monotonically decaying in the direction of cosmological expansion and approaches its infimum only in the limit that the matter sources be asymptotically `turned off'. In the limit of such `extinguished' matter, one may compute $\frac{d^{2}L}{dt^{2}}=0$ and $\frac{d^{3}L}{dt^{3}}=2n\tau(\tau^{2}-\frac{2n\Lambda}{n-1})^{\frac{n}{2}-1}\int_{M}|\partial_{t}K^{tr}|^{2}\mu_{g}<0$ unless $\partial_{t}K^{tr}=0$ as well. 
At this limit, one may simplify the evolution and constraint equations by substituting $K^{tr}=\partial_{t}K^{tr}=0$ to obtain the background warped product spacetimes
\begin{eqnarray}
\label{eq:conf}
\hat{g}\\\nonumber
=-\frac{n^{2}}{(\tau^{2}-\frac{2n\Lambda}{n-1})^{2}}d\tau\otimes d\tau+\frac{n}{(n-1)(\tau^{2}-\frac{2n\Lambda}{n-1})}\gamma_{ij}dx^{i}\otimes dx^{j},\tau\in(-\infty,-\frac{2n\Lambda}{n-1}),
\end{eqnarray}
with $Ric(\gamma)=-\frac{1}{n}\gamma$. Each of these spacetimes admits a globally defined future directed time like conformal Killing field $Y$ given by $Y:=Y^{\mu}\partial_{\mu}=\left(\tau^{2}-\frac{2n\Lambda}{n-1}\right)^{1/2}\partial_{\tau}$ with $L_{Y}\hat{g}=\frac{-2\tau}{(\tau^{2}-\frac{2n\Lambda}{n-1})^{1/2}}\hat{g}$. Therefore we designate these spacetimes as `conformal' spacetimes. Notice that if one turns off the cosmological constant, one retrieves the so called `Lorentz cone' spacetimes analyzed in \cite{fischer2002hamiltonian}
\begin{eqnarray}
\label{eq:milne}
\hat{g}=-\frac{n^{2}}{\tau^{4}}d\tau\otimes d\tau
+\frac{n}{(n-1)\tau^{2}}\gamma_{ij}dx^{i}\otimes dx^{j},\tau\in(-\infty,0),
\end{eqnarray}
which in the `$3+1$' dimensional case is also known as the 'Milne' spacetime. In this special case $Y$ reduces to the \textit{homothetic} Killing field $Y=-\tau\partial_{\tau}$ with $L_{Y}\hat{g}=2\hat{g}$.
The vital question is whether the Lyapunov functional $L$ ever attains its infimum. In an expanding universe model, if the matter sources do not re-collapse to form a singularity, then the matter density falls off and, in the limiting case, may be considered to be negligible. Observing the monotonic decay of the Lyapunov functional along a solution curve, one may be tempted to conjecture the asymptotic stability of a matter filled spacetime. However, we remind the reader again that such a property only provides a weak notion of stability if the matter sources do not develop singularities. Even a vacuum spacetime might be able to go singular whereby the pure gravitational energy density can collapse to form a singularity before the spatial volume of the universe reaches infinity. These questions are related to the \textit{Cosmic Censorship} conjecture and are part of active research.

\section{Perfect Fluids}
A physical cosmology is often built on the assumption that a perfect fluid is the source for Einstein's equations. It is well known that perfect fluids possess the pathological property of forming shock singularities in finite time (such a result was established by D. Christodoulou \cite{christodoulou2007formation} for the case of a perfect fluid evolving on the Minkowski spacetimes). However, we will not address such issues here but rather will observe in a later section that on a sufficiently rapidly expanding spacetime one may avoid shocks by imposing a certain smallness condition on the data. 
The perfect fluid stress-energy tensor
\begin{eqnarray}
\mathbf{T}^{\mu\nu}&=&(P+\rho)u^{\mu}u^{\nu}+P\hat{g}^{\mu\nu}
\end{eqnarray}
 yields
\begin{eqnarray}
E=\mathbf{T}(\mathbf{n},\mathbf{n})=(P+\rho)(Nu^{0})^{2}-P\geq0,\\
S_{ij}:=\mathbf{T}(\partial_{i},\partial_{j})=(P+\rho)u_{i}u_{j}+Pg_{ij}
\end{eqnarray}
with the equation of state $P=(\gamma_{a}-1)\rho$. Now assuming $1\leq\gamma_{a}\leq2$ and a positive mass energy density $\rho$, we obtain that the pressure satisfies $0\leq P\leq \rho$. We use the $n+1$ split of the velocity $u=v-g(u,\mathbf{n})\mathbf{n}$ to express the entities in terms of the spatial ($M$ tangential) velocity $v$. The energy density and momentum flux density tensor $S_{ij}$ read 
\begin{eqnarray}
E=(P+\rho)(1+g_{ij}v^{i}v^{j})-P,
~S_{ij}=(P+\rho)v_{i}v_{j}+Pg_{ij}.
\end{eqnarray}
The energy conditions that are required by our analysis
\begin{eqnarray}
E\geq 0,~(n-2)E+S\geq 0
\end{eqnarray}
take the forms of the following inequalities in the case of a perfect fluid
\begin{eqnarray}
(P+\rho)g_{ij}v^{i}v^{j}+\rho\geq 0,\\
(n-1)(P+\rho)g_{ij}v^{i}v^{j}+nP+(n-2)\rho\geq 0
\end{eqnarray}
which are trivially satisfied under the assumption of $\rho\geq 0$ and $\gamma_{a}\in [1,2]$. In order to extract more dynamical information from our weak Lyapunov functional, we use a conformal technique similar to \cite{fischer2002hamiltonian}. Using the momentum constraint, the second fundamental form may be written as follows (\ref{eq:2ndf}) 
\begin{eqnarray}
\label{eq:fundsplit}
K^{ij}=K^{trij}+\frac{\tau}{n}g^{ij},
\end{eqnarray}
where $g_{ij}K^{trij}=0$, but $\nabla_{j}K^{trij}=-J^{i}$ under ths CMC condition ($\partial_{i} \tau=0$). We use the conformal transformation as described in \cite {fischer2002hamiltonian, moncrief2019could, ashtekar2015general} only replacing the momenta conjugate to the metric $g$ by the second fundamental form through the inverse of a Legendre transformation $\pi=-\mu_{g}(k-(tr_{g}k)g)$. The conformal transformation reads
\begin{eqnarray}
\label{eq:conformal}
(g_{ij},K^{trij})=(\varphi^{\frac{4}{n-2}}\gamma_{ij}, \varphi^{-\frac{2(n+2)}{n-2}}\kappa^{trij}),
\end{eqnarray} 
where $\gamma$ and $\kappa^{tr}$ are the scale-free fields satisfying $\gamma^{ij}\kappa^{tr}_{ij}=0$ with $\gamma\in \mathcal{M}_{-1}$ and $\varphi: M\to \mathbf{R}_{>0}$. Here $\mathcal{M}_{-1}$ is defined as $\mathcal{M}_{-1}=\{\gamma$ is a Riemannian metric on M $|R(\gamma)=-1\}$. 
In reality, the fully reduced dynamics assumes aan equivalence class lying in the orbit space $\mathcal{M}_{-1}/D_{0}$, $D_{0}$ being the group of diffeomorphisms of $M$ isotopic to the identity. To avoid technical complexities, the calculations will simply be restricted to $\mathcal{M}_{-1}$ as the entities we are interested in (such as $\int_{M}\mu_{\gamma}=\int_{M}\sqrt{\det{\gamma_{ij}}}dx^{1}\wedge dx^{2}...\wedge dx^{n}$) are $D_{0}$ invariant. Note that $K^{tr}$ may be decomposed into a transverse-traceless part (with respect to $g$) and a conformal Killing tensor part
\begin{eqnarray}
\label{eq:tracescale}
K^{trij}&=&K^{TTij}+\varphi^{\frac{-2n}{n-2}}(L_{Y}g-\frac{2}{n}\nabla_{m}Y^{m}g)^{ij},
\end{eqnarray}
where $Y\in\mathfrak{X}(M)$. Under this conformal transformation, the components of the conformal Killing tensor $(L_{g}Y-\frac{2}{n}\nabla_{m}Y^{m}g)$ transform as follows 
\begin{eqnarray}
(L_{g}Y-\frac{2}{n}\nabla_{m}Y^{m}g)^{ij}=\varphi^{\frac{-4}{n-2}}(L_{\gamma}Y-\frac{2}{n}\nabla_{m}Y^{m}\gamma)^{ij}
\end{eqnarray}
yielding a consistent transformation of the transverse-traceless tensor $K^{TT}$, that is, $K^{TTij}=\varphi^{-\frac{2(n+2)}{n-2}}\kappa^{TTij}$. Therefore, we may write 
\begin{eqnarray}
\label{eq:trace}
\kappa^{trij}=\kappa^{TT}+(L_{\gamma}Y-\frac{2}{n}\nabla_{m}Y^{m}\gamma)^{ij}.
\end{eqnarray}
Under this conformal transformation, the Hamiltonian constraint (\ref{eq:HC}) becomes the following semi-linear elliptic equation for the conformal factor 
\begin{eqnarray}
\label{eq:LC}
\Delta_{\gamma}\varphi-\frac{n-2}{4(n-1)}\varphi-\frac{(n-2)}{4(n-1)}\varphi^{\frac{-3n+2}{n-2}}|\kappa^{TT}+(L_{Y}\gamma-\frac{2}{n\mu_{\gamma}}\nabla_{m}Y^{m}\gamma)|^{2}\\\nonumber
-\frac{(n-2)E_{g}}{2(n-1)}\varphi^{\frac{n+2}{n-2}}+\frac{n-2}{4n}(\tau^{2}-\frac{2n\Lambda}{n-1})\varphi^{\frac{n+2}{n-2}}=0,
\end{eqnarray}
where $E_{g}$ is the matter energy density that needs scaling under the conformal transformation (\ref{eq:conformal}).
We will scale the matter degrees of freedom in the following way. First, the momentum constraint after the re-scaling of the traceless second fundamental form reads
 \begin{eqnarray}
\nabla[\gamma]_{j}\kappa^{trij}=-\varphi^{\frac{2(n+2)}{n-2}}J_{g}^{i},
\end{eqnarray} 
where $J_{g}^{i}$ is the momentum density without conformal scaling. Now of course, if one chooses the following scaling for the momentum density (York scaling \cite {choquet2009general})
\begin{eqnarray}
J_{g}^{i}=\varphi^{\frac{-2(n+2)}{n-2}}J_{\gamma}^{i},
\end{eqnarray}
the momentum constraint becomes decoupled from the Lichnerowicz equation (\ref{eq:LC}) i.e., 
\begin{eqnarray}
\label{eq:scalem}
\nabla[\gamma]_{j}\kappa^{trij}=-J_{\gamma}^{i}.
\end{eqnarray}  
Now upon substituting $\kappa^{tr}$ from equation (\ref{eq:trace}) into equation (\ref{eq:scalem}) one arrives at an elliptic equation for $Y$
\begin{eqnarray}
-\Delta_{\gamma}Y^{i}+R[\gamma]^{i}_{m}Y^{m}+(1-\frac{2}{n})\nabla[\gamma]^{i}(\nabla[\gamma]_{m}Y^{m})&=&J[\gamma]^{i}.
\end{eqnarray}
The vector field $Y$ therefore depends solely on the metric $\gamma$ and the matter source $J_{\gamma}$, not on the conformal factor $\varphi$. 
Now, utilizing the normalization condition $g_{\mu\nu}u^{\mu}u^{\nu}=-1$, we obtain 
\begin{eqnarray}
\label{eq:normalization}
-(Nu^{0})^{2}+g_{ij}v^{i}v^{j}=-1.
\end{eqnarray}
Since, the right hand side of this equation is a constant (and therefore conformally invariant), each term on the left hand side may be chosen to be conformally invariant leading to the following scalings of $Nu^{0}$ and $v^{i}$
\begin{eqnarray}
(Nu^{0})_{g}&=&(Nu^{0})_{\gamma},v^{i}_{g}=\varphi^{-\frac{2}{n-2}}v^{i}_{\gamma},
\end{eqnarray}
which, together with the expressions for matter energy and momentum density, yields the following scalings for $P, \rho,$ and $E_{g}$
\begin{eqnarray}
(P+\rho)_{g}&=&\varphi^{-\frac{2(n+1)}{n-2}}(P+\rho)_{\gamma},~E_{g}=\varphi^{-\frac{2(n+1)}{n-2}}E_{\gamma}.
\end{eqnarray}
Notice that these scalings are only valid for a perfect fluid source since we have explicitly made use of the normalization relation (\ref{eq:normalization}). 
Rescaling of the matter energy density $E_{g}$ yields the following Lichnerowicz equation 
\begin{eqnarray}
\Delta_{\gamma}\varphi-\frac{n-2}{4(n-1)}\varphi-\frac{(n-2)}{4(n-1)}\varphi^{\frac{-3n+2}{n-2}}|\kappa^{TT}+(L_{Y}\gamma-\frac{2}{n\mu_{\gamma}}\nabla_{m}Y^{m}\gamma)|^{2}\\\nonumber
-\frac{(n-2)E_{\gamma}}{2(n-1)}\varphi^{-\frac{n}{n-2}}+\frac{n-2}{4n}(\tau^{2}-\frac{2n\Lambda}{n-1})\varphi^{\frac{n+2}{n-2}}=0,\nonumber
\end{eqnarray} 
a unique solution of which may be obtained either by the standard sub and super solution technique of \cite {choquet2009general, Oxford University Press} or by a direct method such as the variational technique of \cite{hebey2008variational}. Note that the exponent of $\varphi$ in the term $\frac{(n-2)E_{\gamma}}{2(n-1)}\varphi^{-\frac{n}{n-2}}$ is crucial in proving the uniqueness and existence of  solutions to the Lichnerowicz equation and therefore, the rescaling of the energy density $E_{g}$ is necessary. A straightforward maximum principle argument, applied to the Lichnerowicz equation, shows that the unique positive solution $\varphi=\varphi(\tau,\gamma, \kappa^{TT}, Y,E_{\gamma})$ satisfies 
\begin{eqnarray}
\label{eq:lowerbound}
 \varphi^{\frac{4}{n-2}}\geq\frac{n}{n-1}\frac{1}{(\tau^{2}-\frac{2n\Lambda}{n-1})}
\end{eqnarray}     
with equality holding everywhere on $M$ iff 
\begin{eqnarray}
\kappa^{tr}=\kappa^{TT}+\{L_{Y}\gamma-\frac{2}{n}\nabla_{m}Y^{m}\gamma\}\equiv0,E_{\gamma}\equiv0
\end{eqnarray}
on $M$. Therefore the asymptotic analysis becomes straightforward as the infimum of the Lyapunov functional is attained precisely when the previous two conditions are met. The Lyapunov functional expressed in terms of the conformal variables reads
\begin{eqnarray}
L(\tau,\gamma,\kappa^{tr},E_{\gamma})=\int_{M}(\tau^{2}-\frac{2n\Lambda}{n-1})^{n/2}
\varphi^{2n/(n-2)}(\tau,\gamma,\kappa^{tr},E_{\gamma})\mu_{\gamma}.
\end{eqnarray}
Utilizing the lower bound (\ref{eq:lowerbound}), the infimum of the Lyapunov functional over the space $(\mathcal{M}_{-1}/D_{0})\times S^{tr}_{2}(M)\times \mathcal{E}$ ($\mathcal{E}$ is the space of re-scaled energy densities, $S^{tr}_{2}(M)$ is the space of symmetric covariant traceless 2-tensors with respect to the metric $\gamma$) may be computed as 
\begin{eqnarray}
\inf_{\mathcal{M}_{-1}/D_{0}\times S^{tr}_{2}(M)\times \mathcal{E}}L(\tau,\gamma,\kappa^{tr},E_{\gamma})\\\nonumber
=\inf_{\mathcal{M}_{-1}/D_{0}} \int_{M}(\tau^{2}-\frac{2n\Lambda}{n-1})^{n/2}\varphi^{2n/(n-2)}(\tau,\gamma,\kappa^{tr}=0,E_{\gamma}=0)\mu_{\gamma},\\\nonumber
=(\frac{n}{n-1})^{n/2}\inf_{\mathcal{M}_{-1}/D_{0}}\int_{M}\mu_{\gamma},\\\nonumber
=(\frac{n}{n-1})^{n/2}\left\{-\sigma(M)\right\}^{n/2},
\end{eqnarray} 
where $\sigma(M)$($<0$) is a topological invariant (the higher dimensional analog of the Euler characteristic of a higher genus surface) of the manifold $M$ (of negative Yamabe type considered here). 

The most interesting case here is that of the physical universe (i.e., the $3+1$ dimensional case). Utilizing Ricci-flow techniques, the $\sigma$ constant (and therefore the infimum of the weak Lyapunov functional) of the most general compact 3-manifold of negative Yamabe type has been computed and as such is given by 
\begin{eqnarray}
|\sigma(M)|&=&(vol_{-1}H)^{2/3},
\end{eqnarray}    
where $vol_{-1}H$ is the volume of the hyperbolic part of $M$ computed with respect to the hyperbolic metric normalized to have scalar curvature $-1$ \cite {anderson2004geometrization, anderson2005canonical}. Therefore, apart from the hyperbolic family of $\mathbf{K}(\pi,1)$ manifolds, the remaining components of $M$ i.e., wormholes ($\mathbb{S}^{2}\times \mathbb{S}^{1}$), spherical space forms ($\mathbb{S}^{3}/\Gamma_{k}$), and the graph manifolds (non-hyperbolic parts of $\mathbf{K}(\pi,1)$ manifolds) do not contribute to the $\sigma$ constant. Hence, they do not contribute to the infimum of the weak Lyapunov functional as well. Since the weak Lyapunov functional is geometrically the rescaled volume of $M$ (\ref{eq:LF}), following its monotonic decay towards an infimum dominated only by the hyperbolic components of the spatial manifold, one is led to the natural conclusion that the Einstein flow, in the presence of suitable matter sources and a positive cosmological constant, drives the universe towards an asymptotic state that is volume dominated by hyperbolic components. In other words, the dynamical mechanism at work within the Einstein flow (with or without matter) drives the physical universe towards an asymptotic state which is characterized by a locally homogeneous and isotropic spatial metric.

\section{Stability Results}
In order to understand the extent to which the weak Lyapunov functional decays to its infimum (or may be obstructed from doing so), it is necessary to study the global properties of the solutions. This is indeed a very difficult open problem of classical general relativity. The recent progress on this so called `large data' problem is only limited to a few highly symmetric cases (e.g., Gowdy spacetimes \cite{moncrief1981global, chrusciel1990strong, isenberg1990asymptotic})  or in lower dimensions (e.g., `2+1' vacuum gravity \cite{anderson1997global} where special techniques have been implemented). The Lyapunov functional (\ref{eq:LF}) may always be used to study large data problems. However, as we have mentioned previously, it is not sufficient to control the desired norm of the data ($g,k$ and matter degrees of freedom) where a local existence theorem may be expected to hold. A solution of the Einstein-matter system may break down via curvature concentration or breakdown of the matter field itself before the Lyapunov functional achieves its infimum. In order to investigate the dynamical behaviour of the matter coupled Einstein equations and conclude to what extent the Lyapunov functional decays, it is necessary to study the stability properties of the solutions on which the later achieves its infimum and the matter source vanishes. 

Before tackling this extremely difficult `large' data problem (which is not even resolved in the case of pure vacuum gravity let alone allowing for an arbitrary matter source), it is natural to ask whether solutions sufficiently close to the solutions ((\ref{eq:conf}) or (\ref{eq:milne})) do indeed remain sufficiently close or converge to these backgrounds.
First we consider the pure vacuum limit. In the pure vacuum case, \cite{andersson2004future} studied the stability of the $3+1$ dimensional Lorentz cone spacetimes (\ref{eq:milne}) utilizing the Bel-Robinson energy and its higher order generalizations (which controlled the required $H^{3}\times H^{2}$ norm of the data $(g,k)$). Later these authors \cite{andersson2011einstein} generalized the results for the case of $n\geq 3$ utilizing a wave equation type energy and (its higher order generalization) which played the role of a Lyapunov functional for sufficiently small data. This indeed proved that the non-isolated fixed points playing the role of a center manifold of the suitably re-scaled Einstein dynamics on manifolds that do admit negative Einstein metrics are actual attractors for sufficiently small perturbations. Following these earlier studies several additional ones have been performed addressing the stability issue on manifolds admitting negative spatial Einstein metrics (hyperbolic manifolds for the $3+1$ dimensional case) including source terms in the case of small data. These include the $\Lambda-$vacuum ($\Lambda>0$) \cite{fajman2015stable, puskar2019}, Klein-Gordon field \cite{wang2019future, fajman2021attractors} (with $\Lambda=0$), dust \cite{fajman2021slowly}, and Vlasov matter \cite{andersson2020nonlinear} cases, in particular. The stability of these special solutions incorporating several matter sources that satisfy the energy condition we prescribe for the analysis to hold true provides some `weak' support for our conclusion.

Here we will study the stability properties of a special type of background solution that is isometric to the solution given in the earlier section ((\ref{eq:conf}) or (\ref{eq:milne})). This is done simply because the Milne model (or its equivalent `conformal' spacetime when $\Lambda>0$) is not compatible with physical observation since it is devoid of any matter content. The solutions that we consider are variants of FLRW solutions where the spatial component is given by a compact negative Einstein space (hyperbolic for the $3+1$ dimensional case). These models are the special solutions of the Einstein field equations coupled to a perfect fluid matter source and are explicitly given as follows 
\begin{eqnarray}
\label{eq:flrw}
^{n+1}g&=&-dt\otimes dt+a^{2}(t)\gamma_{ij}dx^{i}\otimes dx^{j},t\in[0,\infty)\\
\label{eq:flrw1}
R[\gamma]_{ij}&=&-\frac{1}{n}\gamma_{ij},a\sim t~~as~~t\to \infty~~for~~\Lambda=0,\\
a(t)&\sim& \sinh(\alpha t)\sqrt{a^{2}(0)+\frac{1}{2\Lambda}}+a(0)\cosh(\alpha t)~~as~~t\to \infty,~~ \Lambda>0,\\\nonumber 
a(0)&:=&a(t=0), \alpha:=\sqrt{\frac{2\Lambda}{n(n-1)}}.
\end{eqnarray}
In addition, the density and the fluid $n-$velocity satisfy $\rho(t)=\frac{C}{a^{n\gamma_{a}}(t)}, v^{i}=0$ where a barotropic equation of state $P=(\gamma_{a}-1)\rho$, $\gamma_{a}\in(1,2)$, is chosen and $C$ is a constant. One may study the global properties of the sufficiently small perturbations of the suitably re-scaled Einstein-Euler field equations about the background solutions described above. The re-scaled Einstein-Euler equations are generically a non-autonomous dynamical system since the scale factor $a(t)$ appears explicitly in the field equations. This is a vital difference from the pure vacuum case where after a suitable re-scaling by powers of the mean extrinsic curvature $\tau$, one may reduce the field equations to an autonomous form. This lack of autonomous character does not necessarily cause a problem provided that the scale factor satisfies a suitable integrability condition (which holds trivially for the current model).  

Before proceeding to the fully non-linear analysis, it is natural to study the stability of the background solutions (\ref{eq:flrw}-\ref{eq:flrw1}) in the regime of linear perturbation theory. Reference \cite{mondal2020linear} proved the linear stability of this Einstein-Euler flow ($t\mapsto (g(t),k^{tr}(t),\rho(t),v(t))$) in the presence of a positive cosmological constant and showed that the perturbed solutions decay to solutions with constant negative spatial scalar curvature that lie sufficiently close to the background solutions. This was accomplished by utilizing a Hodge decomposition of the fluid's $n-$velocity field and it was observed that the pure rotational (`curl' part) and harmonic (topological) contributions decouple at the linear level. This simplified the analysis and a subsequent energy type argument similar to the ones used by \cite{andersson2011einstein} was employed to conclude the result. 

Motivated by the linear stability result, the second author executed a fully non-linear analysis assuming a certain smallness condition on the initial data (\cite{puskarnonlinear}, in prep.). However, construction of a suitable Lyapunov functional for the complete Einstein-Euler system that controls the required norm of the data (contrary to the re-scaled volume functional which only controls the minimum regularity of $(g,k,\rho,v)$) is not straightforward. This is a consequence of the fact that the Euler equations are not of `diagonal' nature while expressed in our choice of CMCSH gauge\footnote{~CMCSH or constant mean extrinsic curvature spatial harmonic gauge: this choice of gauge makes the map $g_{ij}\mapsto R_{ij}(g)$ elliptic thereby allowing one to cast the Einstein evolution equations into hyperbolic form.}. This problem is circumvented by constructing a Lyapunov functional utilizing D. Christodolou's energy current \cite{christodoulou2007formation,christodoulou2016action}. Utilizing the monotonic decay property of this Lyapunov functional for sufficiently small data, it is shown that the perturbed solutions are globally well posed to the future and moreover that they decay to the nearby solutions with constant negative scalar curvature. This result however required that the adiabatic index should lie in suitable range $(\gamma_{a}\in(1,\frac{n+1}{n})$ i.e., that the perturbations are restricted to lie within the so-called `sound cone').    

In $3+1$ dimensions the fixed points are characterized by the condition that the metric be negative Einstein. Following Mostow rigidity, this automatically implies that the manifold be hyperbolic. Contrary to the situation in $3+1$ dimensions, the fixed points are not isolated in higher dimensions. This is due to the fact that the Einstein moduli space is finite dimensional for $n>3$ (but collapses to a point for $n=3$). Such cases are handled by invoking a `shadow' gauge introduced by \cite{andersson2011einstein}. Here however, we only state the result for the physically interesting $3+1$ dimensional case where such moduli spaces do not appear. The following theorem states the stability and asymptotic properties of the small data perturbations of the special solutions described by (\ref{eq:flrw}-\ref{eq:flrw1})

\textbf{Theorem 1:} \textit{Let $(a^{-2}(t_{0})g_{0},a^{-1}(t_{0})k^{tr}_{0},a^{3\gamma_{a}}(t_{0})\rho_{0},a(t_{0})v_{0})\in B_{\delta}(\gamma,0,C_{\rho},0)\subset H^{s}\times H^{s-1}\times H^{s-1}\times H^{s-1},~s>\frac{3}{2}+2$, $t_{0}\in [0,\infty)$, $\Lambda>0$ be the cosmological constant, and $a(t)$ be the scale factor. Assume that the adiabatic index $\gamma_{a}$ lies in the interval $(1,\frac{4}{3})$. Let $t \mapsto (g(t), k^{tr}(t), \rho(t), v(t))$ be the maximal development of the Cauchy problem for the Einstein-Euler-$\Lambda$ flow in constant mean extrinsic curvature spatial harmonic gauge (CMCSH) with initial data $(g_{0},k^{tr}_{0},\rho_{0},v_{0})$. Then there exists a $\gamma^{\dag}\in \mathcal{M}^{\epsilon}_{-1}$  such that $(a^{-2}(t)g, a^{-1}(t)k^{tr}, a^{3\gamma_{a}}(t)\rho, a(t)v)$ flows toward $(\gamma^{\dag},0, C^{'}_{\rho}, 0)$ in the limit of infinite time, that is, 
\begin{eqnarray}
\lim_{t\to\infty}(a^{-2}(t)g(t),a^{-1}(t)k^{tr}(t),a^{3\gamma_{a}}(t)\rho(t),a(t)v(t))=(\gamma^{\dag},0,C^{'}_{\rho},0).
\end{eqnarray}
Here $\mathcal{M}^{\epsilon}_{-1}$ denotes a sufficiently small neighbourhood of the hyperbolic metric in the space of metrics of constant negative scalar curvature (normalized to be $-1$). $C_{\rho}$ and $C^{'}_{\rho}$ are two constants. The convergence is understood in the strong sense i.e., with respect to the available Sobolev norms. 
}

This result suggests that for a universe with accelerated expansion (such as the one considered here), the pathological property of perfect fluid shock formation may be avoided. This roughly indicates that the concentration of energy by non-linearity is dominated by dispersion caused by rapid expansion (a similar observation was also made by \cite{speck2012nonlinear}). In addition to the fully nonlinear stability of the small data perturbations of the special solutions (\ref{eq:flrw}), one is also interested in the causal geodesic completeness of these spacetimes. However, once the relevant estimates are available, it is rather straightforward to prove the causal geodesic completeness. Using the technique of \cite{andersson2004future}, future completeness of the causal geodesics can be established. The geodesic completeness theorem reads 

\textbf{Theorem 2:} \textit{$\exists \delta>0$ such that for any $(a^{-2}(t_{0})g_{0},a^{-1}(t_{0})k^{tr}_{0},a^{3\gamma_{a}}(t_{0})\rho_{0},a(t_{0})v_{0})\in B_{\delta}(\gamma^{*},0,C_{\rho},0,0,0)\subset H^{s}\times H^{s-1}\times H^{s-1}\times H^{s}\times H^{s-1}$,~$s>\frac{3}{2}+2$, the Cauchy problem for the Einstein-Euler system in constant mean extrinsic curvature (CMC) and spatial harmonic (SH) gauge is globally well posed to the future and the space-time is future complete.}

Let us now briefly describe the physical significance of these results. The FLRW model with constant negative spatial \textit{sectional} curvature is globally homogeneous and isotropic since its spatial manifold is simply the hyperbolic $3-$space $\mathbb{H}^{3}$. This is, as we mentioned in an earlier section, compatible with the cosmological principle. However, a more physically meaningful assumption would be \textit{local} homogeneity and isotropy of the physical universe since the astronomical observations that motivate the cosmological principle are limited to possibly a small fraction of the actual universe. If one constructs variants of the FLRW model by considering compact quotients of $\mathbb{H}^{3}$ (by proper and discrete subgroups of SO$^{+}(1,3)$), then such quotient manifolds would certainly satisfy local homogeneity and isotropy criteria since they are locally indistinguishable from $\mathbb{H}^{3}$. These manifolds have constant negative sectional curvature and non-trivial topology (described by their Betti numbers or singular/de-Rahm cohomologies). Our nonlinear stability results suggest that sufficiently small perturbations about these variants of FLRW models are stable to the future (the expanding direction) and moreover that they decay to nearby solutions with constant negative spatial scalar curvature. While these manifolds do admit hyperbolic metrics (we are restricting attention to the physically relevant $3$ spatial dimensions now), the asymptotic state does not necessarily attain the same spatial metric since the hyperbolic metric has constant negative sectional curvature (not just negative scalar curvature). Nevertheless, the asymptotic solutions still lie in a small neighbourhood of the backgrounds (\ref{eq:flrw}-\ref{eq:flrw1}). This in other words implies a \textit{Lyapunov} stability of the solutions described by (\ref{eq:flrw}-\ref{eq:flrw1}). This stability result seems to be unsatisfactory in view of the information provided by the monotonic decay of the \textit{weak} Lyapunov functional where one would expect an \textit{asymptotic} stability of these background solutions. However, we are not able to claim that the monotonically decaying Lyapunov functional ever attains its infimum even asymptotically. This is indeed an open problem of large data long time existence associated with the Einstein flow and closely related to the \textit{Cosmic Censorship Conjecture}. But if the Lyapunov functional ever did achieve its infimum then the infimum would correspond to the background solution which is characterized by the hyperbolic manifold as its spatial component. On the other hand, using the currently available techniques, we can only establish a Lyapunov stability which does not prove (or disprove) the fact that these background solutions are asymptotically stable. These open issues require careful investigation and should be addressed in future studies.

\section{Appendix: Negative Yamabe Manifolds}
In this section, we shall focus attention on the subset of these 3-manifolds
of so-called negative Yamabe type. By definition these admit no Riemannian
metric $\gamma$ having scalar curvature $R(\gamma)\geq 0$. Within the above setting a closed 3-manifold $M$ is of negative Yamabe type if and only if it lies in one of the
following three mutually exclusive subsets \cite{ashtekar2015general, ricci2}: (1) $M$ is hyperbolizable (that
is admits a hyperbolic metric); (2) $M$ is a non-hyperbolizable $K(\pi,1)$ manifold of non-flat type (the six flat $K(\pi,1)$ manifolds are of zero Yamabe type); and
(3) $M$ has a nontrivial connected sum decomposition (i.e., $M$ is composite)
in which at least one factor is a $K(\pi, 1)$ manifold. In this case the $K(\pi,1)$
factor may be either of flat type or hyperbolizable. The six flat manifolds
comprise by themselves the subset of zero Yamabe type. These admit metrics
having vanishing scalar curvature (the flat ones) but no metrics having strictly
positive scalar curvature. Finally manifolds of positive Yamabe type provide
the complement to the above two sets and include the stand-alone $\mathbb{S}^{3}$, the
spherical space forms $\mathbb{S}^{3}/\Gamma_{i}$, $\mathbb{S}^{2}\times\mathbb{S}^{1}$ and connected sums of the latter two
types (recalling that $M\#\mathbb{S}^{3}\approx M$ for any 3-manifold $M$).

It follows immediately from the form of the Hamiltonian constraint that
any solution of the Einstein field equations with Cauchy surfaces of negative
Yamabe type (i.e., diffeomorphic to a manifold in one of the three subsets
listed above) and strictly non-negative energy density and non-negative cosmological
constant (with either or both allowed to vanish) cannot admit a
maximal hypersurface. Thus such a universe model, if initially expanding,
can only continue to do so (until perhaps developing a singularity) and cannot
cease its expansion and ‘recollapse’.

For such manifolds Yamabe’s theorem \cite{yamabe, yamabe2} guarantees that each smooth
Riemannian metric on M is uniquely, globally conformal to a metric $\gamma$ having
scalar curvature $R(\gamma)=-1$. Thus, in a suitable function space setting \cite{yamabe3}
one can represent the conformal classes of Riemannian metrics on $M$ by the
infinite dimensional submanifold $\mathcal{M}_{-1}(M):=\{\gamma\in \mathcal{M}(M)|R(\gamma)=-1\}$
where $\mathcal{M}(M)$ designates the corresponding space of arbitrary Riemannian
metrics on $M$. The quotient of $\mathcal{M}_{-1}(M)$ by the natural action of $D_{0}(M) = Diff_{0}(M)$,
the connected component of the identity of the group $D^{+}(M) = Diff^{+}(M)$ of
smooth, orientation preserving diffeomorphisms of M, defines an orbit space
(not necessarily a manifold) given by $\mathcal{T}M:=\mathcal{M}_{-1}(M)/D_{0}(M)$. Because of
it’s resemblance to the corresponding Riemannian construction of the actual
Teichm\"uller space $\mathcal{T}\Sigma_{p}$ of a higher genus surface $\Sigma_{p}$ we refer to $\mathcal{T}M$ (informally) as the Teichm\"uller space of conformal structures of $M$. The Teichm\"uller space $T\Sigma_{p}$ of the higher genus surface $\Sigma_{p}$ is diffeomorphic
to $\mathbb{R}^{6p-6}$ hence always a smooth manifold. By contrast $\mathcal{T}M$ may either be a manifold or have orbifold singularities or consist of a stratified union of manifolds representing the different isometry classes of conformal Riemannian
metrics admitted by $M$ (i.e., metrics $\gamma$ with $R(\gamma)=-1$).\\

\address{$^1$ Department of Mathematics, Yale University,\\
$^{2}$ Department of Physics, Yale University\\
$^{3}$ Center of Mathematical Sciences and Applications, Department of Mathematics, Harvard University}

\end{document}